\documentclass{aastex}
\usepackage{epstopdf}
\usepackage{amsmath}
\slugcomment{Revised version \today }
\shorttitle{Rotation-vibration wavenumbers of $^{36}$ArH$^+$ and $^{38}$ArH$^+$}
\shortauthors{Cueto et al.}

\begin{document}

\title{NEW ACCURATE MEASUREMENT OF $^{36}$A\lowercase{r}H$^{+}$
 AND $^{38}$A\lowercase{r}H$^{+}$ RO-VIBRATIONAL TRANSITIONS BY HIGH 
RESOLUTION IR ABSORPTION SPECTROSCOPY}

\author{M. Cueto}
\affil{Molecular Physics Department, Instituto de Estructura de 
la Materia (IEM-CSIC).\\ Serrano 123. E-28006 Madrid, Spain}

\author{J. Cernicharo}
\affil{Department of Astrophysics, CAB. INTA-CSIC.\\ Crta 
Torrej\'on-Ajalvir Km 4, E-28850 Torrej\'on de Ardoz, Madrid, Spain}

\author{M. J. Barlow, B. M. Swinyard\altaffilmark{a}}
\affil{Department of Physics and Astronomy, University 
College London.\\ Gower Street, London WC1E 6BT, UK}

\author{V. J. Herrero, I. Tanarro and J. L. Dom\'enech}
\affil{Molecular Physics Department, Instituto de Estructura de 
la Materia (IEM-CSIC).\\ Serrano 123. E-28006 Madrid, Spain}

\altaffiltext{a}{Space Science and Technology Department, 
Rutherford Appleton Laboratory.\\ Didcot OX11 0QX, UK}

\email{jl.domenech@csic.es}

\begin{abstract}

The protonated Argon ion, $^{36}$ArH$^{+}$, has been identified 
recently in the Crab Nebula (Barlow et al. 2013) from Herschel spectra. Given the 
atmospheric opacity at the frequency of its $J$=1-0 and $J$=2-1 
rotational transitions (617.5 and 1234.6 GHz, respectively), 
 and the current lack of appropriate space observatories after the 
recent end of the Herschel mission, future studies on this molecule 
will rely on mid-infrared observations. We report on accurate 
wavenumber measurements of $^{36}$ArH$^{+}$ and $^{38}$ArH$^{+}$
rotation-vibration transitions in the $v$=1-0 band in the range 4.1-3.7 $\mu$m 
(2450-2715 cm$^{-1}$). The wavenumbers of the $R$(0) transitions of 
the $v$=1-0 band are 2612.50135$\pm$0.00033 and 2610.70177$\pm$0.00042
cm$^{-1}$ ($\pm3\sigma$) for $^{36}$ArH$^{+}$ and $^{38}$ArH$^{+}$,
 respectively. The calculated opacity for a gas 
thermalized at a temperature of 100 K and a linewidth of 1\,km\,s$^{-1}$ 
of the {\it R}(0) 
line is $1.6\times10^{-15}\times N$($^{36}$ArH$^+$).  For column densities 
of $^{36}$ArH$^+$ larger than $1\times 10^{13}$ cm$^{-2}$, 
significant absorption by the 
$R$(0) line can be expected against bright mid-IR sources.

\end{abstract}


\section{Introduction}

Molecular complexes 
containing noble gas atoms have been searched in space 
for long without success, using space platforms and ground based 
observatories. The presence of HeH$^{+}$ in the atmospheres of bright 
stars was suggested by Stecher \& Milligan (1961, 1962) to explain the 
differences between observations and stellar models in the ultraviolet 
(see also, Norton 1964; Harris et al. 2004). The role of HeH$^{+}$ in 
the chemistry of the early universe has been also a subject of debate 
(see, e.g., Lepp \& Shull 1984; Galli \& Palla 1998; Bovino et al. 2011) 
and searches have been conducted to try to detect this molecular ion in 
high redshift objects (Zinchenko et al. 2011) unsuccessfully.

Barlow et al. (2013) have recently reported the serendipitous detection 
of $^{36}$ArH$^{+}$ in the emission spectra of the Crab nebula 
during a search for CO lines. In our planet, the most 
abundant Ar isotope is $^{40}$Ar, produced from the disintegration of 
$^{40}$K through electron capture or positron emission, and from beta 
decay of $^{40}$Ca. However, in space the most abundant isotope is 
$^{36}$Ar, produced by alpha processes during stellar nucleosynthesis. 
The abundances on Earth of $^{38}$Ar, $^{36}$Ar and $^{40}$Ar 
are 0.063\%, 0.337\% and 99.600\%, respectively. However, in the Sun 
84.6\% of Argon is $^{36}$Ar (Lodders 2008) and in giant 
planets the ratio $^{36}$Ar/$^{40}$Ar is 8600 (Cameron 1973). 
The high abundance of the heavy $^{40}$Ar on Earth has 
as a consequence that most laboratory studies of ArH$^{+}$ have 
focused on $^{40}$ArH$^{+}$ and that 
little information is available for $^{36}$ArH$^{+}$, and even 
less for $^{38}$ArH$^{+}$. Barlow et al. (2013) have discovered 
$^{36}$ArH$^+$ using the submillimeter Fourier Transform Spectrometer 
(SPIRE, Griffin et al. 2010) on board the Herschel satellite (Pilbratt 
et al. 2010). The frequency of the $J$=1-0 line of $^{36}$ArH$^{+}$
is 617.5 GHz, for which the atmospheric transmission is rather poor even 
for a site as good as that of ALMA. The $J$=2-1 line occurs at 1.235 THz and its 
observation from the ground is impossible due to telluric 
absorption. Hence, space platforms are best for 
the observation of the pure
rotational lines of this molecule in space. As an 
alternative, after the end of the Herschel mission, 
ro-vibrational transitions of  ArH$^{+}$ isotopologues could be 
observed in absorption against bright background mid-infrared sources, 
such as the galactic center. 

There are a number of high resolution spectroscopic studies on all 
isotopologues of this ion (i.e. with $^{40}$Ar, $^{36}$Ar, 
$^{38}$Ar, H and D) in the literature, both in the infrared and in the 
sub-mm wave region. The work by Odashima et al. (1999) is the last published 
addition of laboratory frequency data, and references to previous 
work can be found there. Remarkably, for $^{36}$ArH$^{+}$ and $^{
38}$ArH$^{+}$ the only direct measurements are six and two, 
respectively, ro-vibrational lines, by Filgueira \& Blom (1988) and by 
Haese and Oka (unpublished results, quoted by Johns (1984)), using 
the natural isotopic abundance of $^{36}$Ar and $^{38}$Ar in both cases.

In this Letter we report on the accurate laboratory measurement of 
nineteen lines of the $v$=1-0 band of $^{36}$ArH$^{+}$ and 
$^{38}$ArH$^{+}$ which have been used to fit all available laboratory data 
of all ArH$^{+}$ isotopologues. The mass independent Dunham 
coefficients have been considerably improved. Accurate wavenumbers are 
provided to help in the search and detection of $^{36}$ArH$^{+}$ 
and other isotopologues of ArH$^{+}$ in space in the mid-infrared.

\section{Experimental details}

The apparatus used in this experiment has been reported 
earlier (Dom\'enech et al. 2013, Tanarro et al. 1994),
and has been recently used 
to confirm the identification of NH$_{3}$D$^{+}$ in space 
(Cernicharo et al. 2013). It is based on an IR difference-frequency 
laser spectrometer, a hollow cathode discharge reactor, and a double 
modulation technique with phase-sensitive detection (Domingo et al. 
1994).

Briefly, 
frequency-tunable IR radiation is generated by mixing the 
output of an Ar$^{+}$ laser with that of a tunable ring dye laser in 
a LiNbO$_{3}$ crystal contained in a temperature controlled oven. The 
wavelength coverage is $\sim$2.2-4.2 $\mu$m, with $\sim$3 MHz linewidth 
and $\sim$5\,$\mu$W power. The Ar$^{+}$ laser is 
locked to the $^{127}$I$_{2}$ a$_{3}$
hyperfine component of the $P$(13) 43-0 transition, known with $\sim$0.1
MHz accuracy (Quinn 2003). The laser frequency has a residual 
frequency jitter $<$1 MHz 
and similar long-term stability. The tunable single mode ring dye laser 
is also frequency stabilized, with commercial stabilization electronics 
(residual jitter $<$3 MHz). Its wavelength is measured with 
a high accuracy (10 MHz -3$\sigma$-) commercial wavemeter (High Finesse WSU10), 
calibrated with the stabilized Ar$^{+}$ laser. The  wavemeter accuracy
limits that of the IR frequency scale.

The discharge is modulated at 3.09 kHz through an audio 
amplifier, a step-up transformer and a 940 $\Omega$ ballast resistor. 
Typical discharge conditions are 250 V rms between electrodes and 375 
mA. The IR beam is amplitude modulated at 23.19 kHz 
with an electro-optic modulator and a polarizer placed in the 
path of the Ar$^{+}$ laser beam. The IR beam is split before the 
absorption cell so that one part is directed to an InSb detector and is 
used for noise reduction and the other one goes through the multipass 
cell 32 times (22.4 m pathlength) and is then 
detected by another InSb detector. To improve 
the sensitivity for the $^{36}$ArH$^{+}$ and $^{38}$ArH$^{+}$
isotopologues, an autobalanced transimpedance amplifier 
(ATA) based on the design of Lindsay et al. (2001) has been built
and added to the set-up. 
%Our 
%version of the circuit takes advantage of the fact that the IR beam is 
%amplitude modulated, so the subtracted output of the differential 
%amplifier can be fed to a lockin-amplifier that controls the gain of the 
%reference channel of the ATA. The signal at the sum frequency is not 
%affected by the $\sim$300 ms time constant of the servo, and 
The ATA output is fed to a 
dual-phase lock-in amplifier synchronized at the sum frequency of both 
modulation signals (26.28 kHz). The detection time constant is 30 ms 
with 12 dB/oct roll off and scans have been made at a rate of 
0.005 cm$^{-1}$\,s$^{-1}$ (about 30 time constants per linewidth) to avoid 
lineshape distortion effects. The ATA halves the 
floor noise level, reducing by a factor of four the 
acquisition time necessary to attain a given signal to noise ratio 
(SNR). For $^{36}$ArH$^{+}$ and $^{38}$ArH$^{+}$, between 
100 and 800 scans have been averaged, to obtain SNR's between 
12 and 50. The width of the scans, 0.05 cm$^{-1}$ (about 8 
linewidths FWHM),  keeps the recording time to a minimum and allows a good 
estimation of the baseline.  A 
symmetric triangle wave is programmed to control the scans,
 so the lines are recorded both 
with increasing and decreasing wavenumber scale. The full dataset of 
frequency and intensity values is frequency-binned and averaged in a 
0.0005 cm$^{-1}$ grid.  The wavemeter is recalibrated with the 
frequency stabilized Ar$^+$ laser after every 50 scans.

The reactor is pumped down to 10$^{-3}$ mbar. A continuous 
flow ($\sim$30\,mg\,min$^{-1}$) of Ar with natural isotopic 
composition (l'Air Liquide, 99.9995\% purity) 
at 0.4 mbar pressure is used to generate the ArH$^{+}$
ions in the discharge. 
Attempts to increase the ArH$^{+}$ IR absorption signal by adding 
H$_{2}$ were unsuccessful, indicating that 
the hydrogen concentration necessary to produce maximum ArH$^{+}$
absorption signals is very small and is thought to 
proceed from the tiny amount of residual water. 
ArH$^{+}$ in glow discharge plasmas is known to be produced 
mainly through the reactions H$_{2}^{+}$+Ar$\to$ArH$^{+}$+H 
($k$=2.1$\times $10$^{-9}$ cm$^{3}$s$^{-1}$) and 
Ar$^{+}$+H$_{2}$$\to$ArH$^{+}$+H ($k$=8.7$\times $10$^{-10}$
cm$^{3}$s$^{-1}$) (Anicich 1993); and to be destroyed mainly through the 
reaction ArH$^{+}$+H$_{2}$$\to$H$_{3}^{+}$+Ar 
($4$=6.3$\times $10$^{-10}$\,cm$^{3}$s$^{-1}$). Recent 
kinetic studies on Ar/H$_{2}$ discharges with relatively large 
proportions of H$_{2}$ show appreciable amounts of ArH$^{+}$ 
(M\'endez et al. 2010; Sode et al. 2013).  Nevertheless, previous spectroscopic studies 
of ArH$^{+}$ show that a fairly small proportion of
H$_2$ (ranging from $\sim$0.1 to
 $\sim$10\,\% in the studies of Haese et al. 1983, Johns 1984, 
Laughlin et al. 1987, Brown et al. 1988 and Filgueira \& Blom 1988)  or no supply 
at all of H$_{2}$ (Brault \& Davis 1982) to the discharge 
gives larger ArH$^{+}$ signals.

\section{Results and Discussion}

We have chosen the $R$(6) line of the $v$=1-0 band of $^{40}$ArH$^{+}$ 
at 2711.4029 cm$^{-1}$ as a reference to optimize the operating 
conditions and check the day-to-day repeatability of the experiment. We 
can observe this line in a single scan with SNR of 1100. 
Its Doppler full width at half maximum (FWHM) is 0.0059-0.0061 
cm$^{-1}$, corresponding to a kinetic temperature of $\sim$380-400 K. 
 From the relative line intensities of $R$(6) 
observed in the $v$=1-0 
and $v$=2-1 bands, the estimated vibrational temperature is $\sim$580K. 

We estimate a density of ArH$^{+}$ in the discharge of 
$\sim$4$\times$10$^{10}$\,cm$^{-3}$, 
derived from the transition dipole moments and Herman-Wallis factors for
$^{40}$ArH$^{+}$ in various vibrational bands given by 
Picqu\'e et al. (2000), the estimated vibrational and rotational 
(kinetic) temperatures and the observed peak absorption (tipically $\sim$0.035)
in the $R$(6) $v$=1-0 line of $^{40}$ArH$^+$. This is a rough estimate, 
since ion signals exhibit day-to-day variations of up to 40\%, 
depending on the discharge behavior. Nonetheless this value is close to 
the total electron densities measured in this type of plasmas 
(M\'endez et al. 2010), suggesting that ArH$^{+}$ is a dominant ion 
in our discharge. 

We have measured fourteen lines of $^{36}$ArH$^{+}$
in the $v$=1-0 
band ($P$(6) to $R$(7)) ranging from 2458.4 to 2729.3 cm$^{-1}$. We have 
scanned 0.05 cm$^{-1}$ wide regions, centered on the frequencies 
calculated from the Dunham parameters of Johns (1984). The predictions
are rather accurate, and the lines were found 
within $\pm2.2\times 10^{-4}$ cm$^{-1}$ of their expected value. 
Typically, 200 scans were averaged. The line center position has been 
derived from a least-squares fit of each line to a Gaussian function. As 
an example, the $R$(0) and $R$(3) lines are shown in Figure 1, together with 
their Gaussian fits. The uncertainty in the line positions is estimated 
as the quadratic sum of the standard error of the line center derived 
from the fit plus the manufacturer stated accuracy of the wavemeter, 
i.e. 3.3 MHz (1$\sigma$). The observed line centers and their estimated 
uncertainties ($1\sigma$) are given in Table 1.

These new data for $^{36}$ArH$^{+}$ have been fitted to 
the frequencies derived from the energy levels calculated with a simple 
Hamiltonian for a vibrating rotor (Herzberg 1989)

\begin{equation}
\begin{array}{rl}
E(v,J)= & \nu_0 +B_v J(J+1)-D_v[J(J+1)]^2\\
        & +H_v[J(J+1)]^3+...
\end{array}
\end{equation}

Despite the relatively low number of measurements, the quality of the 
data allows to determine, with statistical significance, 
five independent parameters (up to the $D$ centrifugal 
distortion constants), with an uncertainty-weighted standard deviation 
$\sigma_{w}$=0.61.
 Recall that $\sigma_{w}$ is close 
to unity for an adequate model and reasonable estimates of the 
experimental uncertainties. The low $\sigma_{w}$ value reflects an 
internal coherence of the frequency better than the uncertainty, as it 
is reasonable to expect. As a further check, with the parameters 
determined in 
this fit, we calculate the frequency of the 1-0 rotational 
transition of $^{36}$ArH$^{+}$ at 617524.4$\pm$1.2 MHz ($\pm3\sigma$),
which 
compares very well with the frequency predicted by the Cologne database 
(M\"uller et al. 2005) of 617525.23$\pm$0.45 MHz ($\pm3\sigma$). 
%Fixing $H_{0}$ to the value obtained 
%from the global fit described later, allows for the determination of 
%the $H_{1}$ constant as well, resulting in a prediction 
%for the 1-0 rotational frequency of 617525.4$\pm$1.6 MHz in even better 
%agreement. Nevertheless, we have preferred to give only the constants 
%consigned in Table 1, since they represent the maximum information that 
%can be obtained from our new data alone. 
As noted in the Introduction, prior to this work, only six 
direct measurements of ro-vibrational frequencies of $^{36}$ArH$^{+}$ 
were available. The 
accuracy of the previous values, estimated by their authors, was 
$\sim$0.001 cm$^{1}$. No measurements of $R$(0) have been reported previously.
\begin{figure}
\epsscale{1}
\plotone{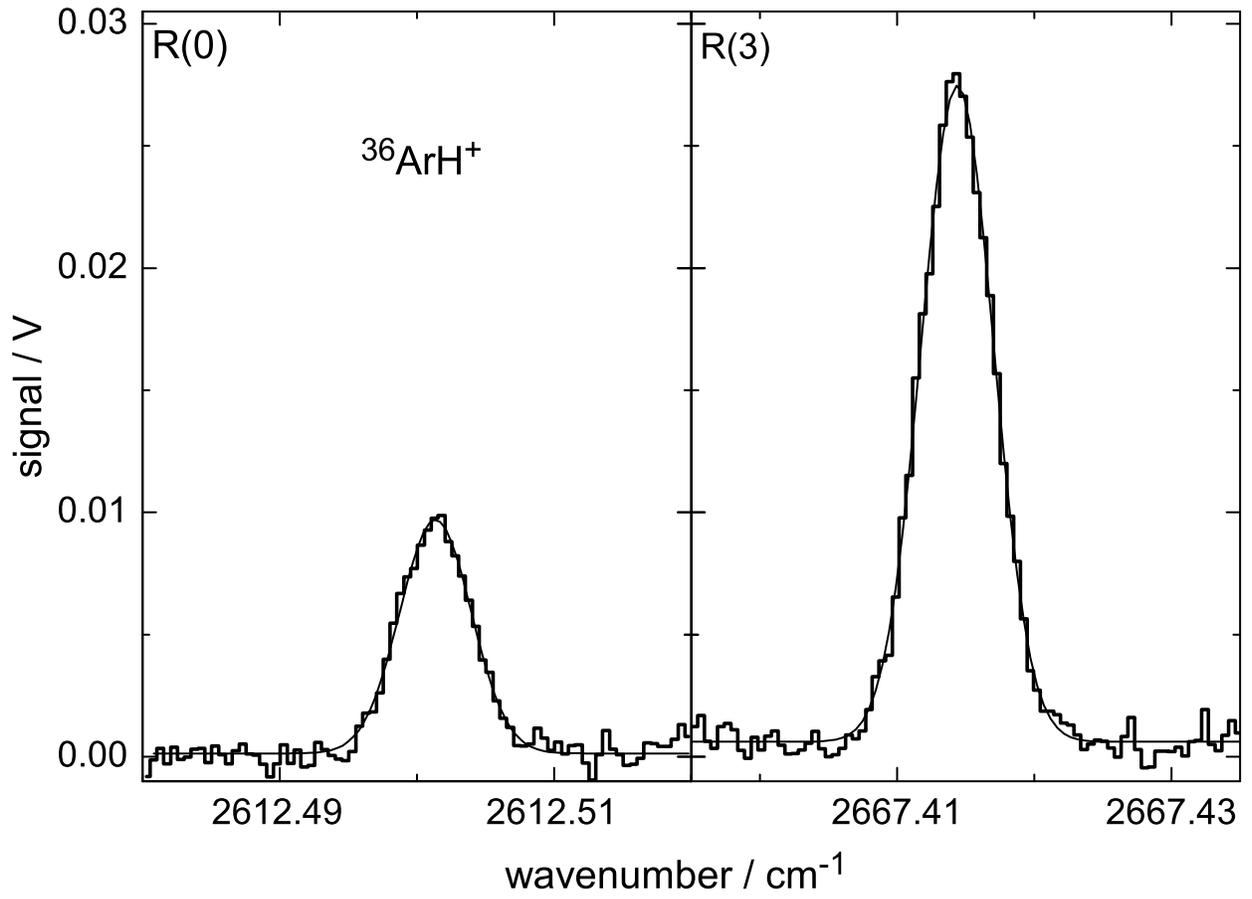}
\caption{$R$(0) and $R$(3) lines of the  $^{36}$ArH$^{+ }$
isotopologue with their Gaussian fits. }
\label{fig1}
\end{figure}

As for $^{38}$ArH$^{+}$, we have recorded five previously unreported lines,
 $R$(0) to $R$(4), in the $v$=1-0 band. Given their weakness,
it was necessary to average up to 800 scans for some of 
them. Since the efficiency of the difference frequency mixing process 
decreases with the IR frequency, and phonon absorption from LiNbO$_{3}$
starts to become significant at lower frequencies, the IR power 
available for lines in the $P$ branch was not sufficient to record them in 
a reasonable amount of time. The wavenumbers determined in this work
and their uncertainties are shown in Table 1.  As an example, the 
$R$(0) and $R$(2) lines are shown in Figure 2, together with their Gaussian 
fits. Filgueira \& Blom (1984) reported the 
observation of the $P$(3) and $P$(4) lines, with $\sim$0.001 cm$^{-1}$ 
accuracy, and, similarly to $^{36}$ArH$^{+}$, no measurements of $R$(0)
have been reported previously.  Given the reduced number of lines 
in this case, we have not 
attempted an independent fit of this isotopologue. 
\begin{figure}
\epsscale{1}
\plotone{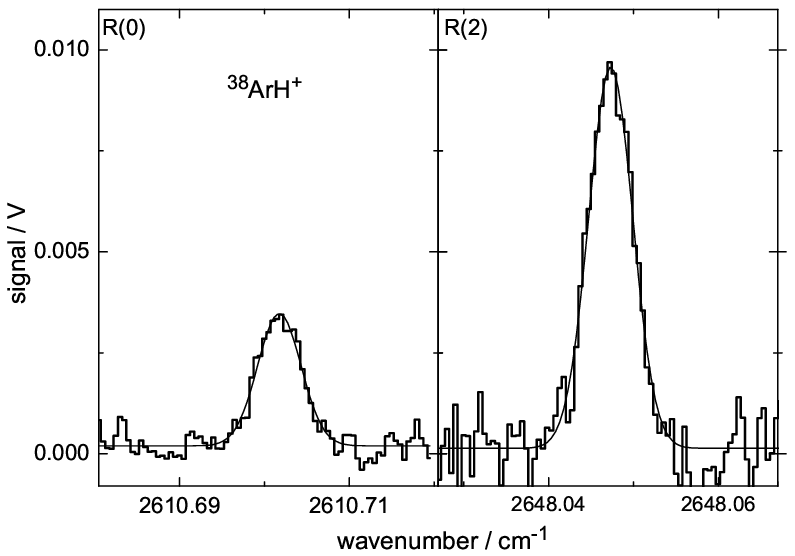}
\caption{$R$(0) and $R$(2) lines of the  $^{38}$ArH$^{+ }$
isotopologue with their Gaussian fits. }
\label{fig2}
\end{figure} 

These new measurements, together with all previous infrared and 
millimeter-wave data found in the literature for all isotopologues of 
ArH$^{+}$ (i.e. those including $^{40}$Ar, $^{36}$Ar, $^{38}$Ar, 
H and D) have been combined in a global fit to a 
mass-independent Dunham's series expansion for a diatomic molecule in a 
$^{1}\Sigma$ state (Watson 1980)

\begin{equation}
\begin{array}{rl}
E(v,J)&=  \sum\limits_{kl}\mu^{-(k/2+l)}U_{kl}\\
 & \times\left(1+m_e\varDelta_{kl}^{\rm{Ar}}/M_{\rm{Ar}}+
        m_e\varDelta_{kl}^{\rm{H}}/M_{\rm{H}}\right) \\
 & \times{\left(v+\frac{1}{2}\right)}^k\left[J(J+1)\right]^l
\end{array}
\end{equation}

\noindent where $U_{kl}$ and $\varDelta^i_{kl}$ are  
mass independent coefficients ($\varDelta^i_{kl}$ are the
Born-Oppenheimer approximation breakdown, or BOB, terms),
 $M_{\rm{Ar}}$ and $M_{\rm{H}}$, 
the atomic masses of the corresponding isotope (Coursey et 
al. 2010), $m_{e}$, the electron mass and $\mu$, 
a charge-modified reduced mass, defined 
for ArH$^{+}$ as

\begin{equation}
\mu=\frac{M_{\rm{Ar}}M_{\rm{H}}} {M_{\rm{Ar}}+M_{\rm{H}}-m_e}.
\end{equation}

In the fitting process it became evident that a significantly 
better global fit is obtained if, in the calculation of $\mu$, 
we use $(M_{\rm{Ar}}-m_e)$ in the 
numerator, instead of the Ar atomic mass $M_{\rm{Ar}}$. In this case, 
$\sigma_{w}$=0.72, 
while that obtained using equation (3) is $\sigma_{w}$=0.86. Using 
$(M_{\rm{Ar}}-m_e)$ amounts to centering 
all the positive charge of the ion on the Ar nucleus. Similar 
improvements in a Born-Oppenheimer potential fit using different 
expressions of $\mu$ were observed in HeH$^{+}$ by Coxon \& 
Hajigeorgiou (1999), although in that case the 
best fit was obtained if half the electron mass was subtracted from that 
of each atom. Furthermore, an ab-initio study of noble gas hydride 
molecule-ions (Schutte 2002) shows a progressive displacement of the 
positive charge of the ion from the H nucleus to the noble gas nucleus, 
at the equilibrium distance, in going from HeH$^{+}$ to KrH$^{+}$.
 Therefore, we deemed advisable to carry out four fits using different 
calculations for the reduced mass: using Watson's (1980) expression; 
splitting equally the charge among both nuclei; completely ignoring the 
electron mass (i.e. treating the molecule as a neutral), and assigning 
all the charge to the Ar nucleus (as described above). The  
$\sigma_{w}$ are 0.86, 0.76, 0.72 and 0.72, respectively. 
In order to better reproduce the 
set of existing observations, we have chosen the latter as our best fit, 
regardless of the possible implications on the adequacy of Equation 
(3) to calculate the charge-corrected reduced mass of 
light ions, or on the effect of $\mu$ on the physical meaning of 
the Dunham parameters thus derived. 

The fit has been carried out using a linear least-squares 
fitting program coupled to the MADEX code (Cernicharo 2012), which 
allows to predict the frequencies of the pure rotational and 
ro-vibrational lines of all the isotopologues of ArH$^+$ 
from the parameters provided by the fitting routine. The fit 
contains 367 experimental frequencies, weighted by the square of their 
estimated reciprocal uncertainty. The results are shown 
in Table 2, together with the parameters obtained by fitting
all the previous data prior to this work.
The parameters with no uncertainty have been fixed to the values derived from
the fitted parameters and the Dunham relations to $U_{k0},U_{k1}$ as provided
by the code ACET of the CPC library written by Ogilvie (1983).  Including a 
BOB term $\varDelta^{\rm{Ar}}_{01}$ does not improve the 
quality of the fit.  
 In both fits,
($M_{\rm{Ar}}-m_e$) has been used in the numerator of Equation (3)
to calculate $\mu$. Summarizing the results, 
before the inclusion of the new data 
reported in this Letter, the standard deviation of the fit was 52.5 MHz, 
which decreases to 50.7 MHz when our new wavenumbers are included. 
The relatively small change in standard deviation is due to the reduced 
number of new accurate data. Nevertheless, significant improvements are obtained  in the 
$U_{10}$, $U_{12}$ and $\varDelta_{10}^{\rm{Ar}}$ parameters,
whose standard deviations decrease by factors of 1.8, 1.4 and 5.6, respectively, when 
the present wavenumbers are taken into account.  The change in the rest of 
parameters 
is marginal, accompanied by a less significative decrease of their standard
deviations, due to
the large number of pure rotational lines measured with high accuracy
for $^{40}$ArH$^+$ and $^{40}$ArD$^+$. The $\varDelta_{01}^{\rm{Ar}}$ BOB parameter is poorly
determined even with those data for the pure rotational lines
(see Odashima et al. 1999).
Using $U_{kl}$ and $\varDelta_{kl}^i$, of Table 2 we can predict
the frequencies of the pure rotational lines of all isotopologues with high
accuracy. The $J$=1-0 transition is predicted at 617525.134$\pm$0.100,
616648.720$\pm$0.100, and 615858.136$\pm$0.100 MHz for $^{36}$ArH$^+$, $^{38}$ArH$^+$, and
$^{40}$ArH$^+$ respectively. The predicted wavenumber of the $v$=1-0 $R$(0) line is
2612.50129$\pm$(6$\times$10$^{-5}$), 2610.70170$\pm$(10$\times$10$^{-5}$), 
and 2609.07719$\pm$(20$\times$10$^{-5}$) cm$^{-1}$ for the same 
isotopologues.
The quoted intervals are $\pm3\sigma$ uncertainties.
Table 3 contains all published 
observed frequencies of  all isotopologues, as well as 
the values calculated from our fit. 
 
The absorption coefficient of the $R$(0) line can be computed from the  
transition dipole moment of the $v$=1-0 transition derived by Picqu\'e et al. (2000). We  
have used their value of $\mu_{1-0}$=0.297\,D based on the calculations
of Rosmus (1979) for the permanent dipole moment of the ground  
vibrational state $\mu_0$=2.2\,D. They also provide a 
 smaller value, $\mu_{1-0}$=0.194\,D, based on the experimental determination
of $\mu_0$=1.42$\pm$0.6\,D by Laughlin et al. (1987). However, 
 Laughlin et al. (1989) provided $\mu_0$=3.0$\pm$0.6\,D after a
refinement of their experiment, in  better agreement
with the value by Rosmus (1979). Both values
are in good agreement with the calculations performed at a higher level of theory  
by Cheng et al. (2007) and by the analysis of the experimental potential
curve by Molski (2001). Using $\mu_{1-0}$=0.297\,D  we can derive
$\tau$($R$(0))=$\alpha$(T)$\times N$(ArH$^+$)/$\varDelta$v, where 
$N$(ArH$^+$) is the column density of the molecule in cm$^{-2}$, $\varDelta$v is
the linewidth in km\,s$^{-1}$ and where $\alpha$(T)
is 9.5$\times$10$^{-15}$, 1.6$\times$10$^{-15}$, 3.2$\times$10$^{-16}$, and 1.5$\times$10$^{-16}$
(in cm$^2$km\,s$^{-1}$), for $T$=10, 100, 500, and 1000\,K, respectively. 
%These values for $\alpha$(T) have to be decreased by a factor $\sim$2.4 if the  
%value for $\mu_0$ of Laughlin et al. (1987) is adopted, or
%increased by a factor $\sim$1.9 if the experimental value of Laughlin et al.  
%(1989) is assumed. 
These values for $\alpha$(T) have to be decreased by a factor $\sim$2.4, 
or increased by a factor $\sim$1.9, depending on wether  the
$\mu_0$ value of Laughlin et al. (1987), or that of Laughlin et al. (1989),
is assumed.  
The best conditions to detect the $R$(0) lines are those of the 
diffuse ISM lines of sight where the
kinetic temperature could be below 100\,K and no emission is expected 
from the $R$(0) line. For the estimated column densities by Barlow et al. (2013)
absorption in the $R$(0) lines of a few percent could be expected. 
We note, however, that if the kinetic temperature of the gas in 
the CRAB nebula is above 1000 K, the $R$(0) line, and others, 
could appear in emission.  For column densities above
10$^{13}$ cm$^{-2}$ the different isotopologues of ArH$^+$ could be easily
detected in cold dark clouds against bright sources 
through mid-IR observations.

\section{Concluding remarks}

We have provided direct accurate wavenumber measurements of nineteen 
vibration-rotation lines of $^{36}$ArH$^{+}$ and 
$^{38}$ArH$^{+}$, in natural isotopic abundance, measured with a difference 
frequency laser spectrometer and a hollow cathode discharge cell. Of 
those, only six had been reported before, and with much larger 
uncertainty. Furthermore, the 
new wavenumbers have improved the Dunham-type fit to 
all published rotation and vibration-rotation data for all isotopologues 
of this molecule, allowing for more accurate predictions of other 
transitions for any of them.
Notably, the measured wavenumbers of the $R$(0) transitions of the 
$v$=1-0 band are 2612.50135$\pm$0.00033 and 2610.70177$\pm$0.00042 cm$^{-1}$ 
($\pm3\sigma$) for $^{36}$ArH$^{+}$ and $^{38}$ArH$^{+}$ 
[predicted values of 2612.50129 $\pm($6$\times$10$^{-5}$) cm$^{-1}$ and
 2610.70170 $\pm$(10$\times$10$^{-5}$) cm$^{-1}$], 
respectively. These wavenumbers should help in future searches for 
absorptions of these molecules against bright sources. 
\acknowledgements

The authors acknowledge the financial support from the Spanish MINECO 
through the Consolider Astromol project, grant CSD2009-00038. JLD and MC 
acknowledge additional support through grant FIS2012-38175. VH and IT 
acknowledge additional support through grant FIS2010-16455. JC 
acknowledge additional support through grants AYA2009-07304 and 
AYA2012-32032. Our skillful technicians J. Rodr\'{i}guez and M.A. Moreno, 
are gratefully acknowledged.

\clearpage

\begin{deluxetable}{crrrrrl}
%%\tabletypesize{\small}
 \tablewidth{0pt}
\tablecaption{Observed line centers, their estimated 1$\sigma$ uncertainties, 
and spectroscopic constants of $^{36}$ArH$^+$.}
\tablenum{1}
\tablehead{
 \colhead{Isotopologue}
&\colhead{Line}
&\colhead{$\nu_{\rm{obs}}$/cm$^{-1}$}
& \colhead{$\sigma$ \tablenotemark{a}}
& \colhead{(o-c) \tablenotemark{b}}
& \colhead{}
& \colhead{Constant/cm$^{-1}$ \tablenotemark{c}}}

\startdata
 & $P$(6) & 2458.36336 & 11.4 & 0.3 & $B_{0}$ & 10.30044364(778)\\
 & $P$(5) & 2482.47613 & 11.3 & -5.5 & $D_{0}$ & 6.21374(154)$\times $10$^{-4}$ \\
 & $P$(4) & 2505.91727 & 10.5 & 7.7 & $\nu_{1}$ & 2592.651339(42) \\
 & $P$(3) & 2528.67068 & 11.6 & 3.8 & $B_{1}$ & 9.92620133(616) \\
 & $P$(2) & 2550.72091 & 11.8 & -8.9 & $D_{1}$ & 6.127689(908)$\times $10$^{-4}$ \\
 & $P$(1) & 2572.05291 & 13.2 & -2.8 & & \\
$^{36}$ArH$^+$ & $R$(0) & 2612.50135 & 11.3 & 5.9 & & \\
 & $R$(1) & 2631.58798 & 11.1 & -10.6 & & \\
 & $R$(2) & 2649.89731 & 10.3 & 8.6 & & \\
 & $R$(3) & 2667.41441 & 10.4 & -0.3 & & \\
 & $R$(4) & 2684.12561 & 11.9 & 4.6 & & \\
 & $R$(5) & 2700.01671 & 11.4 & -8.8 & & \\
 & $R$(6) & 2715.07445 & 11.3 & 0.9 & & \\
 & $R$(7) & 2729.28504 & 11.0 & 2.0 & & \\
\hline 
 & $R$(0) & 2610.70177 & 13.9 & & & \\
 & $R$(1) & 2629.76268 & 11.2 & & & \\
 $^{38}$ArH$^{+}$  & $R$(2) & 2648.04731 & 13.4 & & & \\
 & $R$(3) & 2665.54197 & 14.9 & & & \\
 & $R$(4) & 2682.23225 & 13.9 & & & 
\enddata

\tablenotetext{a}{$\sigma$= estimated uncertainty / 10$^{-5}$\,cm$^{-1}$}
\tablenotetext{b}{(o-c)=($\nu_{\rm{obs}}-\nu_{\rm{calc}}$) / 10$^{-5}$\,cm$^{-1}$}
\tablenotetext{c}{Numbers in parentheses are one standard deviation in units of the last 
quoted digit, as derived from the fit.}
\end{deluxetable}

\begin{deluxetable}{ccrr}
\tablewidth{0pt}
\tablecaption{Mass independent Dunham coefficients\tablenotemark{+} for ArH$^+$}
\tablenum{2}
\tablehead{\colhead{$k$} & \colhead{$l$} & \colhead{$U_{kl}$ (this work)\tablenotemark{a}} 
&\colhead{$U_{kl}$ (previous data)\tablenotemark{a}}}
 
\startdata
1 & 0 &      2688.29968(193)                      & 2688.29896(345)                    \\ 
2 & 0 &       -60.56901(224)                      &  -60.56869(227)                    \\
3 & 0 &         4.9333(124)$\times$10$^{-01}$     &    4.9320(123)$\times$10$^{-01}$   \\
4 & 0 &        -2.348(287)$\times$10$^{-03}$      &   -2.332(284)$\times$10$^{-03}$    \\
5 & 0 &        -1.154(316)$\times$10$^{-04}$      &   -1.162(313)$\times$10$^{-04}$    \\
6 & 0 &        -1.330(132)$\times$10$^{-05}$      &   -1.329(131)$\times$10$^{-05}$    \\
0 & 1 &        10.28307619(216)                   &   10.28307575(219)                 \\
1 & 1 &        -3.6901057(708)$\times$10$^{-01}$  &   -3.6900543(738)$\times$10$^{-01}$\\
2 & 1 &         2.92102(466)$\times$10$^{-03}$    &    2.91688(478)$\times$10$^{-03}$  \\
3 & 1 &         9.25(165)$\times$10$^{-06}$       &    1.028(168)$\times$10$^{-05}$    \\
4 & 1 &        -2.898(275)$\times$10$^{-06}$      &   -3.022(279)$\times$10$^{-06}$    \\
5 & 1 &        -1.948(165)$\times$10$^{-07}$      &   -1.896(167)$\times$10$^{-07}$    \\
0 & 2 &        -6.0181686(568)$\times$10$^{-04}$  &   -6.0181497(568)$\times$10$^{-04}$\\
1 & 2 &         7.89698(914)$\times$10$^{-06}$    &    7.8769(126)$\times$10$^{-06}$   \\ 
2 & 2 &         1.469(381)$\times$10$^{-08}$      &    2.146(410)$\times$10$^{-08}$    \\
3 & 2 &        -2.1555(349)$\times$10$^{-08}$     &   -2.2871(351)$\times$10$^{-08}$   \\ 
4 & 2 &         2.887511467(0)$\times$10$^{-10}$  &    4.188353696(0)$\times$10$^{-10}$\\
5 & 2 &        -2.276400996(0)$\times$10$^{-11}$  &   -2.531285219(0)$\times$10$^{-11}$\\
0 & 3 &         1.54073(178)$\times$10$^{-08}$    &    1.54146(187)$\times$10$^{-08}$  \\  
1 & 3 &        -3.867(101)$\times$10$^{-10}$      &   -3.745(128)$\times$10$^{-10}$    \\
2 & 3 &        -1.441(182)$\times$10$^{-11}$      &   -1.732(210)$\times$10$^{-11}$    \\
3 & 3 &         2.367548148(0)$\times$10$^{-13}$  &    6.839849573(0)$\times$10$^{-13}$\\
4 & 3 &        -9.706972163(0)$\times$10$^{-13}$  &   -1.029987077(0)$\times$10$^{-12}$\\
0 & 4 &        -8.961(261)$\times$10$^{-13}$      &   -9.148(293)$\times$10$^{-13}$    \\
1 & 4 &         1.074791089(0)$\times$10$^{-15}$  &    2.300842563(0)$\times$10$^{-15}$\\
2 & 4 &        -1.091227090(0)$\times$10$^{-14}$  &   -1.135430007(0)$\times$10$^{-14}$\\
3 & 4 &        -4.145809861(0)$\times$10$^{-16}$  &   -4.269036526(0)$\times$10$^{-16}$\\
4 & 4 &         3.101458722(0)$\times$10$^{-16}$  &    3.303414639(0)$\times$10$^{-16}$\\
0 & 5 &         2.203620896(0)$\times$10$^{-17}$  &    2.202997632(0)$\times$10$^{-17}$\\
1 & 5 &        -3.102378289(0)$\times$10$^{-18}$  &   -3.367869218(0)$\times$10$^{-18}$\\
2 & 5 &         1.657426695(0)$\times$10$^{-18}$  &    1.894902940(0)$\times$10$^{-18}$\\
3 & 5 &        -5.622789166(0)$\times$10$^{-19}$  &   -5.889490755(0)$\times$10$^{-19}$\\
0 & 6 &        -2.567213307(0)$\times$10$^{-21}$  &   -2.537578088(0)$\times$10$^{-21}$\\
1 & 6 &        -1.281992858(0)$\times$10$^{-21}$  &   -1.307682397(0)$\times$10$^{-21}$\\
2 & 6 &        -4.750748688(0)$\times$10$^{-22}$  &   -5.139236544(0)$\times$10$^{-22}$\\
3 & 6 &         2.274040168(0)$\times$10$^{-22}$  &    2.408609282(0)$\times$10$^{-22}$\\
0 & 7 &        -3.390838692(0)$\times$10$^{-26}$  &   -3.984572007(0)$\times$10$^{-26}$\\
1 & 7 &         1.797128568(0)$\times$10$^{-25}$  &    1.996397008(0)$\times$10$^{-25}$\\
2 & 7 &        -4.768691261(0)$\times$10$^{-26}$  &   -4.646346036(0)$\times$10$^{-26}$    
\enddata
\tablenotetext{a}{Units are cm$^{-1}$amu$^{k/2+l}$.  Numbers in parentheses are
$1\sigma$ uncertainties in units of the last quoted digit.  
An uncertainty of 0 means that the parameter has been fixed.}
\tablenotetext{+}{The following BOB 
parameters (dimensionless) have also been fitted 
(this work[from the fit to the previous data]): \\
$\varDelta_{10}^{\rm Ar}$= 1.824(145)$\times$10$^{-01}$  $[$2.017(813)$\times$10$^{-01}$$]$\\    
$\varDelta_{10}^{\rm H} $=-3.3110(122)$\times$10$^{-01}$ $[$-3.3126(125)$\times$10$^{-01}$$]$\\   
$\varDelta_{20}^{\rm H} $= 6.147(218)$\times$10$^{-01}$  $[$6.157(221)$\times$10$^{-01}$$]$\\     
$\varDelta_{01}^{\rm H} $= 1.25816(439)$\times$10$^{-01}$ $[$1.25312(498)$\times$10$^{-01}$$]$\\ 
$\varDelta_{11}^{\rm H} $= 8.006(279)$\times$10$^{-01}$  $[$7.854(290)$\times$10$^{-01}$$]$\\    
$\varDelta_{02}^{\rm H} $= 1.0098(384)                  $[$9.929(399)$\times$10$^{-01}$$]$\\    
}
\end{deluxetable}

\begin{deluxetable}{c r r r r l r r r r r c}
%\rotate
\tablewidth{0pt}
\tabletypesize{\scriptsize}
\tablecolumns{12}
\tablecaption{Measured and calculated ro-vibrational transitions of the isotopologues of ArH$^+$}
\tablenum{3}
\tablehead{
\colhead{Isotopologue}  &   \colhead{$J_{\rm u}$} & \colhead{$J_{\rm l}$} & \colhead{$v_{\rm u}$} & \colhead{$v_{\rm l}$} & 
\colhead{Units} &   \colhead{$\nu_{\rm obs}$}   & \colhead{$\sigma_{\rm obs}$} & \colhead{$\nu_{\rm cal}$}  & 
\colhead{$\sigma_{\rm cal}$}  &  \colhead{$\nu_{\rm obs}$-$\nu_{\rm cal}$} & \colhead{Ref.\tablenotemark{1}} 
}
\startdata
 $^{36}$ArH$^+$     &    7 &  6 &  1 &  0 &cm$^{-1}$&     2715.0745 &    .00011 &       2715.074333&   .000034&   .000117& 1 \\
 $^{36}$ArH$^+$     &    8 &  7 &  1 &  0 &cm$^{-1}$&     2729.2850 &    .00011 &       2729.285196&   .000041&  -.000156& 1 \\
 $^{36}$ArH$^+$     &    4 &  3 &  1 &  0 &cm$^{-1}$&     2667.4140 &    .00100 &       2667.414412&   .000021&  -.000412& 2 \\
 $^{36}$ArH$^+$     &    3 &  2 &  1 &  0 &cm$^{-1}$&     2649.8960 &    .00100 &       2649.897267&   .000022&  -.001267& 2 \\
 $^{36}$ArH$^+$     &    2 &  1 &  1 &  0 &cm$^{-1}$&     2631.5890 &    .00100 &       2631.588125&   .000024&   .000875& 2 \\
 $^{36}$ArH$^+$     &    1 &  2 &  1 &  0 &cm$^{-1}$&     2550.7200 &    .00100 &       2550.720959&   .000026&  -.000959& 2 \\
 $^{36}$ArH$^+$     &    2 &  3 &  1 &  0 &cm$^{-1}$&     2528.6720 &    .00100 &       2528.670669&   .000024&   .001331& 2 \\
 $^{36}$ArH$^+$     &    3 &  4 &  1 &  0 &cm$^{-1}$&     2505.9180 &    .00200 &       2505.917300&   .000022&   .000700& 2 \\
 $^{36}$ArD$^+$     &    1 &  0 &  0 &  0 &MHz      &   319065.3790 &    .06500 &     319065.389737&   .005378&  -.010737& 3 \\                    
 $^{38}$ArH$^+$     &    1 &  0 &  1 &  0 &cm$^{-1}$&     2610.7018 &    .00014 &       2610.701706&   .000036&   .000064& 1 \\
 $^{38}$ArH$^+$     &    2 &  1 &  1 &  0 &cm$^{-1}$&     2629.7627 &    .00011 &       2629.762543&   .000034&   .000137& 1 
\enddata
\tablecomments{Table 3 is published in its entirety in the electronic edition
the Journal.  A portion is shown here for guidance regarding its form and content.  
Frequencies are in cm$^{-1}$ or in MHz as indicated. These units apply to all the
elements of each row. Observed and predicted uncertainties ($\sigma_{\rm obs}$ and $\sigma_{\rm cal}$) 
are 1$\sigma$ values. Transitions with $\sigma_{\rm obs}$=0 have been discarded in the fit due to
a large $\nu_{\rm{obs}}-\nu_{\rm{cal}}$ value.}
\tablenotetext{1}{References. --- (1) This work; 
(2) Filgueira \& Blom (1988);
(3) Bowman et al. (1983);(4) Brown et al. (1988);
(5)  Liu et al. (1987); (6) Brault \& Davis (1982);
(7) Johns (1984); (8) Laughlin et al. (1988);
(9) Odashima et al. (1999)}        
  
\end{deluxetable}

\end{document}